\documentclass[final,authoryear,5p,times,twocolumn]{elsarticle}
\usepackage{natbib}
\usepackage{graphicx}
\usepackage{latexsym}
\usepackage{epsfig} 
\usepackage{amssymb,amsbsy}
\usepackage{amsmath}
\usepackage{eucal}
\usepackage{ulem}
\usepackage{longtable}
\usepackage{xtab,booktabs}
\usepackage[dvipsnames]{color}

\newcommand{\be}{\begin{equation}}
\newcommand{\ee}{\end{equation}}
\def\ba{\begin{eqnarray}}
\def\ea{\end{eqnarray}}

\def\msun{M_\odot}

\def\ltsima{$\; \buildrel < \over \sim \;$}
\def\simlt{\lower.5ex\hbox{\ltsima}}
\def\gtsima{$\; \buildrel > \over \sim \;$}
\def\simgt{\lower.5ex\hbox{\gtsima}}
\def\fr{\frac}

\newcommand{\pol}[1]{\stackrel{\rm LCP}{\mathrm{RCP}}}

\newcommand{\me}{\mathrm{e}}

\renewcommand{\a}{\alpha}

\newcommand{\s}{\sigma}

\def\apj{{\it Astrophys.~J.}}
\def\aj{{\it Astronom.~J.}}
\def\apjl{{\it Astrophys.~J.~Lett.}}

\def\mnras{{\it Mon.~Not.~R.~Astron.~Soc.}}
\def\nature{{\it Nature}}
\def\science{{\it Science}}
\def\icarus{{\it Icarus}}
\def\aap{{\it Astron.~Astrophys.}}

\def\apss{{\it Astrophys. Space Sci.}}

\journal{}
 
\begin{document}

\begin{frontmatter}
\title{Age Aspects of Habitability}

\author{M.~Safonova,~J.~Murthy}
\address{Indian Institute of Astrophysics, Bangalore, India}
\ead{rita@iiap.res.in, murthy@iiap.res.in}
\author{Yu.~A.~Shchekinov}
\address{Department of Space Physics, SFU, Rostov on Don, Russia}
\ead{yus@sfedu.ru}

\begin{abstract} 
\noindent

A ‘habitable zone’ of a star is defined as a range of orbits within which a rocky planet can support
liquid water on its surface. The most intriguing question driving the search for habitable planets is
whether they host life. But is the age of the planet important for its habitability? If we define habitability as
the ability of a planet to beget life, then probably it is not. After all, life on Earth has developed within
only ~800 Myr after its formation --- the carbon isotope change detected in the oldest rocks indicates the
existence of already active life at least 3.8 Gyr ago. If, however, we define habitability as our ability to detect
life on the surface of exoplanets, then age becomes a crucial parameter. Only after life had evolved
sufficiently complex to change its environment on a planetary scale, can we detect it remotely through its
imprint on the atmosphere --- the so-called biosignatures, out of which the photosynthetic oxygen is the most
prominent indicator of developed (complex) life as we know it. Thus, photosynthesis is a powerful biogenic
engine that is known to have changed our planet’s global atmospheric properties. The importance of
planetary age for the detectability of life as we know it follows from the fact that this primary process,
photosynthesis, is endothermic with an activation energy higher than temperatures in habitable zones, and is
sensitive to the particular thermal conditions of the planet. Therefore, the onset of photosynthesis on planets
in habitable zones may take much longer time than the planetary age. The knowledge of the age of a planet is
necessary for developing a strategy to search for exoplanets carrying complex (developed) life --- many
confirmed potentially habitable planets are too young (orbiting Population I stars) and may not have had
enough time to develop and/or sustain detectable life. In the last decade, many planets orbiting old (9--13
Gyr) metal-poor Population II stars have been discovered. Such planets had had enough time to develop
necessary chains of chemical reactions and may carry detectable life if located in a habitable zone. These old
planets should be primary targets in search for the extraterrestrial life.

\end{abstract}

\end{frontmatter}

\section{Introduction}
\label{sec:1}

Habitability may be quantitatively defined as a measure of the ability of a planet to develop and sustain 
life \citep{schu}; its maximum is set as 1 for a planet where life as we know it has formed, thus it is 1 for the Earth. 
The requirement for a planet to be called habitable (or potentially habitable)\footnote{Both definitions {\it habitable} and
{\it potentially habitable} are used in the literature, meaning essentially the same, but see Sec.~4 for our 
discussion on the definition.} is that the planet is located within the host's HZ and has terrestrial characteristics: 
rocky, with a mass range of 0.1--10 Earth masses and a radius range of $0.5\sim$ 2 Earth radii\footnote{The latest 
simulations have shown that after $\sim$1.7 Earth radii the planets are of increasingly lower density, indicating that 
they are less rocky and more like mini-Neptunes, placing the Earth's twin limit on the radius (for ex. Buchhave et al. 2014), 
though uncertainties remain, see, e.g. Torres et al. (2015)}. 
A habitable zone (HZ) is conservatively defined as a region where a planet can support liquid water on the surface (Huang 1959). 
The concept of an HZ is, however, a constantly evolving one, and many different variations of it have been since 
suggested (see, for example, an excellent review by Lammer et al. (2009) and references therein, and Heller \& 
Armstrong (2014a) as a more recent one). Biogenic elements (such as C, H, N, O, P and S) have also 
been considered as necessary complementary factors for habitability \citep{3c}, but their presence is implied by 
the existence of water as they are produced in the same stars \citep{wo,jp}.

We would like to stress here that throughout the paper, when we talk about detecting life on exoplanets, we still 
mean life as we know it, the presence of which we are able to establish through predictable changes in planetary 
atmospheres. Even on Earth, there is a possibility of a different kind of life not based on a usual triad -- 
DNA--protein--lipid, see, for example, discussion on a ``shadow" biosphere by Davies et al. (2009). 
But just as on Earth we are not able to find it (yet) as we do not know `where or what to look for', 
we may not be able to distinguish these different kinds of life from the natural environments
of exoplanets. Hence, when we talk about biosignatures, we mean only biosignatures
that our kind of life produces --- oxygen, ozone, nitrous oxide, etc (e.g. Seager et al. 2012).  A planet may host life as 
we know it (in other words, be not just {\it habitable} but {\it inhabited}), but we will still not detect it unless it has 
evolved sufficiently to change its environment on a planetary scale, for instance, through the production of an 
oxygen atmosphere 
by photosynthetic organisms. Photosynthesis is currently the only geologically documented biogenic process 
(see e.g. Lyons and Reinhard, 2011; Fomina and Biel, 2014 and references therein) that can provide sufficient energy to 
modify the global planetary (or atmospheric) properties. The large free energy release per electron transfer and stability of 
the oxygen molecule due to its strong bonding ensures that an oxygen-rich atmosphere provides the largest feasible energy source 
for compex life (e.g. Catling et al. 
2005). Therefore, by analogy with the
Earth, we presume the presence of an oxygen atmosphere as necessary for a planet to host a complex life. Such life would have
modified the global planetary
 (or atmospheric) properties to be noticed from 
space, and from very far away; after all, the closest potentially habitable planet is at about 12 light years ($\tau$ Ceti) 
and we cannot go there to verify. Even Mars might still be inhabited by a primitive subsurface biota which are 
undetectable without a local and detailed examination. It may also be possible for life to evolve in a manner that we 
have not anticipated, which, even if it changes the environment globally, would not be detectable simply because we 
are not looking for those particular changes. For example, aphotic life can exist in the subsurface oceans of Europa 
or Enceladus, but such life would be currently impossible for us to detect {\it ex situ}. 

Biological methanogenesis was suggested as a rival to the
photosynthesis process in changing the global environment
and capable of enriching the exo-atmospheres with biogenic
methane (Schindler \& Kasting 2000; Kharecha et al. 2005).
Kharecha et al. (2005) has shown that the rate of biogenic
methanogenesis in the atmosphere of an Archaen Earth
could have been high enough to enrich the atmosphere with
high concentration of biogenic methane. However, planets
with reduced mantles might enrich their atmospheres by methane abiotically (e.g., Etiope \& Lollar 2013), 
and thus methane alone cannot guarantee habitability. From this point
of view, methanogenic products are a less certain biosignature
of Earth-like life than oxygen (Seager et al. 2012). Accounting
for a competitive interrelation between metabolic and abiotic
origins of methane, a more conservative understanding suggests that only the simultaneous presence of methane along
with other biogases is a reliable indication of life (e.g., Selsis
et al. 2002; Kaltenegger et al. 2007; Kiang et al. 2007;
Kasting et al. 2014). It could also be that the planet never de-
velops oxygenic photosynthetic life. In such cases, other biomarkers have been suggested; for example, dimethyl
disulphide and CH3Cl may be detected in infrared (IR) in the
planetary atmospheres of low-ultraviolet (UV) output stars
(Domagal-Goldman et al. 2011).

Carter (1983) has pointed out that the timescale for the evolution of intelligence on the Earth ($\sim 5$ Gyr) is 
comparable to the main sequence lifetime of the Sun ($\sim 10$ Gyr). Lin et al. (2014) suggested that intelligent life 
on expolanets can be detected through the pollution it inflicts on the atmosphere. However, intelligent life, once evolved, 
is no longer in need of a very precisely defined biosphere --- we can already create our own biospheric habitats on planets 
that are lifeless in our definition of habitability, for ex. Moon or Mars, though we are intelligent for only a 0.0000026\% of the 
time life exists on Earth: 100 years out of 3.6 Gyr. Therefore, intelligent life may not be so easily detectable, especially if they 
had longer time to evolve. However, to answer the most important question of ``are we alone", we do not necessarily need to 
find intelligent life. Even detection of a primitive life will have a profound impact on our civilization. Therefore, 
we need to concentrate on the period in the planet’s history when the emerged life had already influenced the atmosphere 
of the planet in a way that we can possibly recognize.  

We discuss here the importance of the age of the planet in the
evaluation of whether that HZ planet contains life and whether
that life is detectable. We examine the plausibility of a discovery
of a habitable planet with detectable biota among the close
(within 600 pc) neighbours of the Sun. We argue that variations
in their albedos, orbits, diameters and other crucial parameters
make the formation of a significant oxygen atmosphere take
longer that the current planetary age and thus, life can be detectable on only half of the confirmed PHPs with a known age.

\section{Initial stages of habitability} 

Necessary conditions for the developing of life are thought to include rocky surface and liquid water; however, the 
aspects connected with the stages preceding the onset of biological era are usually left out of consideration. 
Planetary age as a necessary condition for life to emerge was first stressed by \citet{hua} and implicitly 
mentioned by \cite{cri73} in their concept of a Directed Panspermia.  

In order to understand the importance of planetary age for
the evolution of a detectable biosphere we will consider, as an
example, the development of cyanobacteria and the related atmospheric oxidation
 \citep{irw}. This process involves several endothermic reactions and requires 
sufficiently high temperatures to be activated. In general, the temperature dependence of the photosynthetic 
rate is rather complicated and conditionally sensitive, with the effective activation energy being of the order of 
tens of kJ mol$^{-1}$ \citep{aoxy}, much higher than the typical equilibrium temperature on habitable 
planets. Thus small variations in atmospheric and crust properties can considerably inhibit photosynthesis 
and increase the growth time of the mass of cyanobacteria. This conclusion may be illustrated through the consideration
of the elementary process of carboxylation of RuBP (ribulose-1,5-bisphosphate: C$_5$H$_{12}$C$_{11}$P$_2$) in the 
dark Benson-Bassham-Calvin cycle of photosynthesis \citep{calvin,farq}. These photosynthetic reactions, controlled by 
enzymes, are known to be very sensitive to ambient temperatures with an optimum rate at around 40$^\circ$C, 
and a practically zero rate outside the temperature range of $0^{\circ}<t<60^{\circ}$C \citep{toole}. Amongst 
other fundamental factors RuBP carboxylation is probably the most relevant one, determining the optimal temperature of 
photosynthesis, and is characterized by the activation energy $V_{\rm C}\simeq 30-60$ kJ~mol$^{-1}$ at the growth 
temperature \citep{aoxy}. We can roughly characterize the RuBP carboxylation by the Arrhenius law 
\be
k_{\rm C}=A\me^{-V_{\rm C}/T},
\ee 
where $k_{\rm C}$ is the rate constant, $A$ the prefactor, and $T$ is the  absolute temperature. 
The characteristic time of RuBP carboxylation is $\tau_{\rm C} \propto k_{\rm C}^{-1}\propto\exp(V_{\rm C}/T)$. 
Since the RuBP carboxylation is one of the main processes optimizing photosynthetic reactions, $\tau_{\rm C}$ can roughly
characterize the rate of photosynthesis on a planet. 

The range of variation in $\tau_{\rm C}$  
on a habitable planet due to the uncertainty in the equilibrium temperature $T_e$ is 
\be
\fr{|\delta\tau|}{\tau_{e}}=\fr{V_{\rm C}}{T_e}\fr{|\delta T|}{T_e}\,,
\ee
with $\tau_{e}$ being a characteristic time of photosynthesis at $T_e$.  The equilibrium temperature $T_e$, 
in turn, is calculated using planetary parameters inferred from the observations,
\be
T_e=\left[\fr{L (1-a)}{16\pi\sigma\varepsilon r^2}\right]^{1/4}\,,
\ee
where the uncertainties in the parameters determine the uncertainty in its estimate,
\be\label{vari}
\fr{|\delta T|}{T_e}=\fr{1}{4}\left(\fr{|\delta L|}{L}+\fr{|\delta a|}{1-a}+|\delta\varepsilon|+2\fr{|\delta r|}{r}\right)\,.
\ee
Here $L$ is the luminosity of the central star, $a$ and $\varepsilon$ the planet's albedo and emissivity, $r$ the 
orbital radius, and $\sigma$ the Stephan-Boltzmann constant. It is readily seen that the actual time of the onset of
photosynthesis for a given 
habitable planet might differ significantly from the value calculated from largely uncertain parameters that were, in turn, 
derived from observables. Indeed, uncertainties in estimates of the equilibrium temperature $|\delta T|/T_e$ are heavily amplified 
for habitable planets with $V_{\rm C}/T_e\simeq 10-20$ for $V_{\rm C}\simeq 30-60$ kJ~mol$^{-1}$ and $T_e\sim$300 K, 
such that even relatively low observational errors in deriving the parameters in Eq.~(\ref{vari}), say 5\% each, might result in 
$50-100$\% error in the estimates of the overall photosynthesis rate. If one considers oxidation of the Earth atmosphere as a 
process tracing the developing photosynthesis, the characteristic time for the growth of biota on early Earth can 
be estimated as the oxidation time, $\tau_{\rm O_2}\sim 2$ Gyr (Kasting 1993, Wille et al. 2007, Fomina and Biel, 2014). 
Therefore, a 50\% error in $\tau_{\rm C}$ may delay the possible onset of biological evolution on a planet by 1 Gyr, i.e. 
biogenesis might not start earlier than 3 Gyr from the planetary formation. In general, however, 
the problem of photosynthetic process is much more complex, depending on many factors determined by thermal and 
non-thermal processes on a planet \citep{aoxy,oxy}, and might be even more sensitive to variations in physical conditions. 
Even on the early Earth, physical conditions could have been such as to preclude the onset of biogenesis over a long time 
\citep{sag, mah, sol}.

From this point of view,  the planetary habitability index (PHI) recently proposed by \cite{schu} in the form
\be
\text{PHI}_0=(S\cdot E\cdot C \cdot L)^{1/4},
\label{eq:5}
\ee
can be generalized with explicit inclusion of the age of the planet $t$ as
\be\label{eq:6}
\text{PHI}(t)=\text{PHI}_0\prod\limits_{i}^{}(1-\me^{-t/t_i}).
\ee
In Eq.~(\ref{eq:5}) $S$ defines a stable substrate, $E$ the necessary energy supply, $C$ the polymeric chemistry and $L$ the liquid 
medium; all the variables here are in general vectors, while the corresponding scalars represent the norms of these vectors. In
Eq.~(\ref{eq:6}), the index $i$ denotes a chemical chain relevant for further biochemical evolution, and $t_i$ is its characteristic time. 
It is obvious that the asymptotic behaviour -- approaching the maximum habitability -- is controlled by the slowest process with 
the longest $t_i$.

\section{Other Factors Delaying the Onset of Habitability}

Sagan (1974) was the first to stress that harmful endogenous
and exogeneous processes in the early Earth could postpone
emergence of life on it. Such processes could be important
even in the very initial primitive episodes of biogenesis and
delay the formation of biota for up to billions of years.
 It is known from the $^{182}$W isotope dating that the late heavy bombardment 
(LHB) of Earth, Moon and Mars lasted till about 3.8 Ga \citep{{scho},{moy},{lb}}. The Martian primitive atmosphere is 
believed to have been lost through catastrophic impacts about 4 Ga (e.g., Melosh et al. 1989, Webster et al. 2013). 
Evidence of a heavy bombardment in other exoplanet systems exists: collision-induced hot dust was detected in several 
young planetary systems. Spectral signatures of warm water- and carbon-rich dust in the HZ of a young $\sim 1.4$ Ga MS star 
$\eta$ Corvi \citep{lat}, and of host dust in seven sun-like stars (Wyatt et al. 2007) indicate recent frequent catastrophic 
collisions between asteroids, planetesimals or even possible planets (Song et al. 2005). Out of these seven stars, five are 
young systems within their first Gyr of life. 

It is also well-known that solar-type stars remain very active in the first billion years of their life, sustaining 
conditions that are hostile to the survival of the atmosphere and to the planetary habitability.  
G-type stars, within the first 100 Myr of reaching ZAMS, produce continuous flares of EUV radiation up to 100 times 
more intense than the present Sun, and have much denser and faster stellar winds with an average wind density
of up to 1000 times higher. Low-mass K- and M-type stars remain X-ray and EUV-active longer than solar-type stars,
where EUV emission can be up to 3--4 times and 10--100 times, respectively, higher than G-type stars of the same age; and 
active M-type stars could keep stellar winds in the HZ that are at least 10 times stronger than that 
of present Sun (France et al. 2013). 

In recent simulations by Schaefer et al. (2014) of the development of oceans on super-Earths, it was shown 
that though larger than the Earth planets keep their oceans for longer (up to 10 Gyr),  it takes longer for them
to develop the surface ocean due to the delayed start of volcanic outgassing that returns water back to the surface 
from the mantle. For super-Earths 5 times the Earth's mass, that would take about a billion years longer.  The 
oceans are believed to have established on Earth 750 Myr after formation, therefore super-Earths would have
their surface water established at only 2 Gyr after formation. After all, the Great Oxygenation Event
about 2.5 Ga (Anbar et al. 2007) was most likely induced by
oceanic cyanobacteria, which allowed life to emerge on land
about 480 -- 360 Ma (Myr ago) (Kenrick and Crane, 1997).

\section{Potentially Habitable Planets}

At the time of writing, more than 1900 exoplanets have been
confirmed (Extrasolar Planets Encyclopaedia, June 2015)
with another 4000 waiting for confirmation (NASA
Exoplanet Archive). The majority of detected planets is in the 
vicinity of the Sun, and their hosts are mostly young Population~I (Pop~I) stars with ages of hundreds of Myr 
to a few Gyr. The age distribution of the host stars with measured ages is shown in Figure 1. $58\%$ of the 
host stars have ages of 4.5 Gyr and less, and more than a third ($\sim 38\%$) are younger than 3 Gyr. Simple statistics 
shows the median age of $\sim 3.8$ Gyr. 

\begin{figure}[h!]
%\begin{center}
%\hskip -0.4in
\includegraphics[scale=0.4]{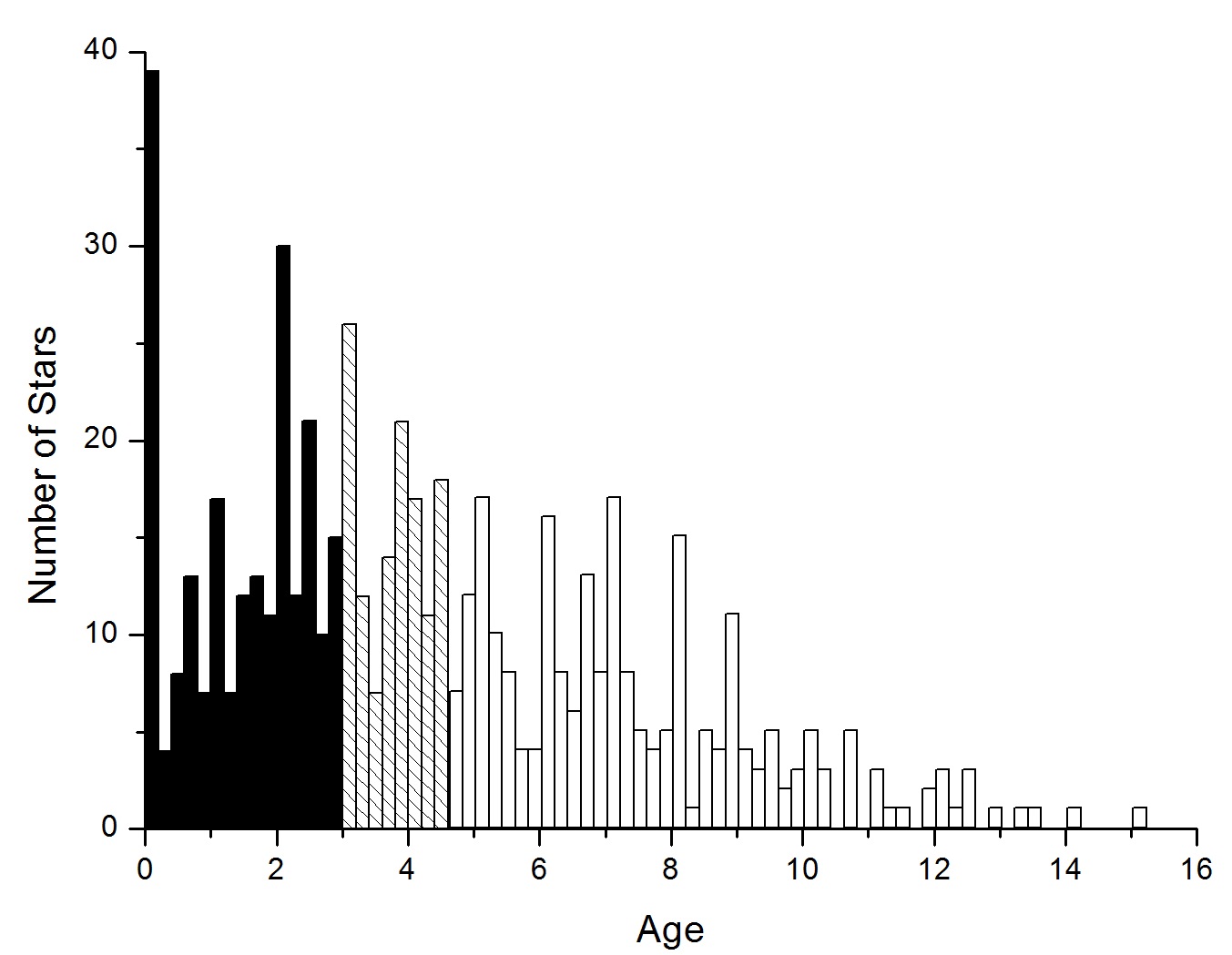}
\caption{Age distribution of the stars hosting confirmed planets (total 583 hosts with known ages at time of writing). 
We highlight the number of stars with ages below 3 Gyr in black, and between 3 and 4.5 Gyr as hatched.
The predominance of young host stars is clearly seen, which could be the effect of observational selection \citep{ssm}. 
This figure was made using the Extrasolar Planets Encyclopaedia data.}
\label{fig:example}
%\end{center}
\end{figure} 

The fact that more than a third of the planetary systems, discovered by ongoing exoplanetary 
missions, are younger than $\sim 3$ Gyr is not surprising, because the continuous star formation (SF) in the Galactic disk 
supplies young stars, and the fraction of hosts younger than 3 Gyr represents that very fraction of Pop~I stars that 
would be born provided the star formation rate is nearly constant during the whole period of the thin disk formation. 
Most of current exoplanet missions suffer from an observational bias --- they mostly detect systems that are younger 
than the age at which life is presumed to have appeared on the Earth.\footnote{The earliest geological/fossilized
evidence for the existence of biota on Earth dates to $\sim$ $3.8-3.5$ Ga, see Brack et al. (2010) and 
references therein.}

Incidentally, Fig.~1 shows a deficit of stars with ages $t>6$ Gyr. Assuming that Pop~I stars, 
i.e. the thin Galactic disk, have started forming at around 10 Ga \citep{ch03,ca07}, one might expect the 
presence of such old stars in our vicinity in the proportion corresponding to the SF history in the early Galaxy. 
The most conservative assumption implies a constant SF rate, in which case one should expect the number of
planet-hosting stars with ages $t>6$ Gyr of about 40\%. It is, however, believed that 
the star formation was more active in the early epochs \citep{bou}, therefore, the fraction of hosts older than 
6 Gyr should be 
correspondingly higher. The reason for the decline in the number of the hosts in this age range is unclear and might, in 
particular, indicate that planetary systems lose planets with age.

About 40 potentially habitable planets (PHPs) are currently documented\footnote{See, for example, the online 
Habitable Exoplanets Catalog (HEC), maintained by the Planetary Habitability Laboratory at the University 
of Puerto Rico, Arecibo, {\tt http://phl.upr.edu/projects/habitable-exoplanets-catalog}, but not exclusively.}, 
though extrapolation of Kepler's data shows that in our Galaxy alone there could 
be as many as 40 billion PHPs (Petigura et al.  2013). In Table~\ref{table:ages} we show the data for PHPs for which the 
host ages were available in the literature. The fraction of young planetary systems is nearly consistent with the 
age distribution of Pop~I stars: among the 33 confirmed habitable planets with known ages more than half are 
$\lesssim 3.5$ Gyr old. 

It seems reasonable to update the definitions in footnote~1 on page~2 as
\begin{enumerate} 
\item \emph{Potentially Habitable planet} -- a rocky, terrestrial-size planet in a HZ of a star.
\item \emph{Habitable planet} --  a rocky, terrestrial-size planet in an HZ
with detected surface water and some of the biogenic gases
in atmosphere.
\item \emph{Inhabited planet} -- the best case scenario: a rocky, terrestrial-size planet in a HZ with simultaneous 
detection of species such as water, ozone, oxygen, nitrous oxide or methane in atmosphere, as proposed by e.g. 
Sagan et al. (1993) or Selsis et al. (2002). 
\end{enumerate}
We may expect only a primitive form of biota on the youngest planets ($\lesssim 2$ Gyr) in Table~1, which would not
be detectable. Biogenesis could have started, or even progressed to more advanced stages with an oxidized 
atmosphere, on older planets with ages from 2 to 4 Gyr. In the former case, one can 
expect that methane from metabolic reactions has already filled the atmosphere, while in the latter case, oxygen 
molecules at some level can appear in the atmosphere ---  though atmospheric oxygen on Earth appeared about 2.5 Ga,
the Earth itself became visibly habitable only about 750--600 Ma, when the biosphere became active 
and complex enough to modify the environment to be noticed from space (e.g. Mend\'{e}z et al. 2013). 
The traces of these gases may, in principle, be observed in sub-mm and micron 
wavelengths, provided the planets are orbiting low-mass stars (0.5--0.8 $ \msun$). Even if a third of the low-mass 
stars in the sky host planets \citep{tut}, there may be as many as a thousand planets within a 10 pc vicinity with ages 
ranging from Myr to a few Gyr.

\begin{table*}[h!]
\begin{center}
\caption{Host ages for {\it confirmed} potentially habitable planets.}
\begin{tabular}{llcccl}
\hline
 Star	           & Planet(s)	                      &  Age Estimate                           & Metallicity                & Distance             & Ref. to age\\
                 &                                        & ~(Gyr)~                                    & [Fe/H]                   &(pc)               	&            \\\hline
Kepler 61    & Kepler-61 b$^{\dagger}$        &    $\sim 1$                                &  $0.03$                  & 326      		&  1       \\
Gliese 667C & Gl 667 c                               &  $<2$; 2--5; $>2$                     & $-0.59$                   & 7.24         		&  1; 2; 3   \\
Kepler 62    & Kepler-62 e,f                         & $7\pm 4$                                & $ -0.37 $                 & 368   			 &  4         \\
Kapteyn's    & Kapteyn's b                          &  10--12                                   & $-0.99$                   & 3.91   			  &  5         \\  
Gliese 163   & Gl 163 c                               &$3.0 (+7., -2.)$; $>2$; $6\pm 5$ & $0.1$                      & 15       		&  1; 7; 8   \\
HD 40307    & HD 40307 g                         & $1.2\pm 0.2$; 4.5; 6.1               & $ -0.31$                  & 12.8   			  &  9; 10 \\  
HD 85512    & HD 85512 b                         & $5.61\pm 0.61$                        & $-0.33$                   & 11      			 &  11         \\     
Kepler 22     & Kepler-22 b$^{\dagger}$       &   $\sim 4$                                & $ -0.29$                  & 190    			  &  12         \\
Gliese 832   & GJ 832 c                               &  9.24                                       & $ -0.31\pm 0.2$         & 4.95    			 &  13 \\
Kepler 186   & Kepler-186 f                         &  $4 \pm 0.6$                            & $ -0.26\pm 0.12 $     & $\sim$172     	&  14         \\
Kepler 296   & Kepler-296 e,f$^{\dagger}$   &   4.2 (+3.4, -1.6)                       & $-0.12\pm 0.12$       &  $\sim$226 		  &  14  \\
Kepler 436   & Kepler-436 b$^{\dagger}$     & 3.0 (+7.7, -0.3)                          &     $0.01\pm 0.1$     &  $\sim$618     	&   14        \\
Kepler 437   & Kepler-437 b$^{\dagger}$     & 2.9 (+7.5, -0.3)                          &$0.00\pm 0.1$          &   $\sim$417  	  &   14        \\
Kepler 438   & Kepler-438 b                        & 4.4 (+0.8, -0.7)                         &   $0.16\pm 0.14$     &    $\sim$145    	 &   14        \\
Kepler 439   & Kepler-439 b$^{\dagger}$     & 7.2 (+3.6, -3.9)                          &      $0.02\pm 0.1$    & $\sim$693   	  &   14        \\
Kepler 440   & Kepler-440 b$^{\dagger}$     & 1.3 (+0.6, -0.2)                         &    $-0.3\pm 0.15$      &$\sim$261		 &   14         \\
Kepler 441   & Kepler-441 b                        & 1.9 (+0.65, -0.4)                       &   $-0.57\pm 0.18$     &$\sim$284 			 &    14         \\
Kepler 442   & Kepler-442 b                        & 2.9 (+8.1, -0.2)                         & $-0.37\pm 0.1$        &$\sim$342 			 &    14         \\
Kepler 443   & Kepler-443 b$^{\dagger}$     & 3.2 (+7.5, -0.4)                          &  $-0.01\pm 0.1$       & $\sim$779  			&    14         \\
KOI 4427     & KOI 4427 b$^{\dagger}$     &    3.6 (+2.6, -1.3)                       &    $-0.07\pm 0.14$    & $\sim$240  		 &  14      \\         
Kepler 174   & Kepler-174 d$^{\dagger}$     &      $7.\pm 4.$                           & $-0.556$                 &  360   				  & 15 \\
Kepler 309   & Kepler-309 c$^{\dagger}$     &         1.5                                   & $0.415$                   &  581   			&    16  \\
Kepler 421   & Kepler-421 b                        &         $4.\pm 0.8$                     & $-0.25$                    &  320   				&  15  \\
Kepler 108   & Kepler-108 c                        &         $8.9\pm 3.7$                    & $-0.026$                  & 861   				&  15 \\
Kepler 397   & Kepler-397 c                        &         $0.6\pm 3.8$                    & $-0.035$                   &  1154   			&  15 \\
Kepler 90     & Kepler-90 h                          &         $0.53\pm 0.88$                & $-0.17$                    & 835   				&  15 \\
Kepler 87     & Kepler-87 c                          &         $0.5\pm 3.7$                   & $-0.17$                     &  782 				 &  15 \\
Kepler 69     & Kepler-69 c                          &         $0.4\pm 4.7$                   & $-0.29$                     &  360   			&  15 \\ 
Kepler 235   & Kepler-235 e$^{\dagger}$      &        1.5                                   & $0.087$                   &  525 				 &  16  \\
Kepler 283   & Kepler-283 c$^{\dagger}$    &   2.0                                       & $-0.26$                     &  534.4   			&    16 \\
Kepler 298   & Kepler-298 d$^{\dagger}$    &  1.5                                         & $ -0.121$                  & 474.3			      &   16   \\
 EPIC 201367065& EPIC 201367065 d          &  $2\pm 1$                               & $ -0.32\pm 0.13$       & 45                       & 17 \\
tau Ceti      & tau Ceti e                              & 5.8                                         & $-0.55\pm 0.05$        &  3.65                     & 18 \\
\hline
\end{tabular}
\label{table:ages}
\end{center}
\vskip -0.1in
$^{\dagger}$ The radii of these planets are $>1.7$ Earth's, however, it is still too soon to exclude them from the list, according
to Torres et al. 2015, since there are many uncertaintites in the modelling of the transition from rocky to
hydrogen/helium planets, and these planets may be rocky.\\[0.05in]
References to ages: 1. The Extrasolar Planet Encyclopaedia ({\tt http://exoplanet.eu}); 2. Angala-Escud\'{e} et al. 2012; 
3. Anglada-Escud{\'e} et al. 2013; 4. Borucki et al. 2013; 5. Anglada-Escud{\'e} et al. 2014; 6. Mamajek \& Hillenbrand 2008; 
7. Tuomi \& Anglada-Escud{\'e} 2013; 8. Open Exoplanet Catalogue ({\tt http://www.openexoplanetcatalogue.com}); 
9. Nordstr{\"o}m et al. 2004; 10. Tuomi et al. 2013a; 11. Pepe et al. 2011; 
12. Metcalfe 2013; 13. Wittenmyer et al. 2014; 14. Torres et al. 2015; 15. NASA Exoplanet Archive at 
{\tt http://exoplanetarchive.ipac.caltech.edu}; 16. Gaidos 2013; 17. Crossfield et al. 2015; 18. Tuomi et al. 2013b.
\end{table*}

The age of a planet is of primary importance for developing
the future strategy of looking for life on PHPs. Since space programs are extremely expensive and require extensive valuable
telescope time, it is crucial to know in advance which planets
are more likely to host detectable life. 
 Young planets will not have atmospheres abundant in products of photosynthetic processes,
and many planets, though residing in the HZ, may not actually be habitable for life as we know it. For example, the host stars in the 
Degenerate Objects around Degenerate Objects (DODO) direct imaging search for sub-solar mass objects around white 
dwarfs (Hogan et al. 2009) are rather young with an average age of only 2.25 Gyr. The target star selection of the Darwin 
(ESA) mission is restricted to stars within $10-25$ pc (Kaltenegger \& Fridlund 2005), and two space missions that are 
currently under study, the NASA Transiting Exoplanet Survey Satellite (TESS) mission and ESA’s PLAnetary Transits and 
Oscillations of stars (PLATO) mission, will only survey bright F, G, K stars and M stars within 50 pc (e.g. Lammer et al. 2013), 
sampling therefore only the thin Galactic disk stars --- young Pop~I hosts. The main focus of Exoplanet 
Characterization Observatory (EChO) \citep{echo} is the observation of hot Jupiter and hot Neptune planets, limited due 
to the mission lifetime constraints to bright nearby M stars (Tinetti et al. 2012). Most known habitable planets cannot have an 
existing complex biosphere although they may develop it in the future, because most currently known PHPs are found around 
relatively young Pop~I stars. We feel reasonable to fix a period of $\sim 4$ Gyr as the minimum necessary time for the formation 
of complex life forms at optimal conditions, as evidenced by the Earth's biosphere. Direct observations of planetary atmospheres
 in IR and sub-mm wavebands would be a promising method for tracing biogenesis. Planned future IR and sub-mm observatories 
could provide such observations (see discussion in Sec.~\ref{sec:prospects} below.)

In this context, we undertaken the project of updating the catalog of Nearby Habitable Systems (HabCat) constructed for SETI by 
Turnbull and Tarter (2003a) for the search for potentially habitable hosts for complex life. A complete characterization of all 
the stars within a few hundred (or even a few tens 
of) parsecs, including their masses, ages, and whether they have planetary systems (including
terrestrial planets), was not realizable at that time. Our aim was to find out the information on 
these stars: their ages and whether they have planets and if they could be potentially habitable.

To begin with, we have taken the HabCat~II, a ``Near 100'' subset -- a list of the nearest 100 star systems of 
the original HabCat \citep{habcat2}, as a basis for our project. These stars were scrutinized for information on 
their age, nearby planets etc., which were missing in the original catalogue but are important now due to their impact 
on selecting the targets for future space missions. Out of 100 nearby (within 10 pc) objects in the HabCat~II, we have found the
 age data for 50 stars. This list is being cross-correlated with the Hypatia Catalog, which is a project to find abundances for 
50 elements, specifically bio-essential elements, for the stars in the HabCat (Hinkel et al. 2014). Our goal is to compile a list of
the most probable planets that may allow future missions to
search our neighbourhood for habitable/inhabited planets
more efficiently. The preliminary result of this project is presented in Table A.2 in Appendix B.

\section{Old Planetary Systems}

\subsection{General census}

Most of the old planetary systems were discovered serendipitously. Only in 2009 were targeted surveys of metal-poor 
stars initiated (Setiawan et al. 2010). In spite of that, quite a few old ($\gtrsim 9$ Gyr) planetary systems are currently known.
 \cite{ssm} attempted to compile a 
list of such system (see their Table~1) on the basis of metallicity, considering stars $[Fe/H] \leq -0.6$. They, however, missed many 
previously known systems with ages determined by several different methods, including metallicity abundances, 
chromospheric activity, rotation and isochrones. Combining their table with other studies (Saffe et al. 2005; Haywood 2008 and 
latest updates of online expolanet catalogs) brings the census of planetary systems 
with ages $\gtrsim 9$ Gyr to 116 planets (90 host stars; see Appendix~A for the table of these systems). It is possible 
that the number of such hosts is much larger since we have counted only those stars where estimates from different methods were 
comparable. For example, in the list of NASA Exoplanet Archive candidates to PHPs, out of 62 hosts with 
estimated ages, 28 are older than 10 Gyr. 

The majority of old planets was detected by the radial velocity (RV) method which is biased to detect preferentially massive 
planets due to a limited sensitivity. Continuously increasing precision of radial-velocity surveys may in future change this 
picture, and the first example of that is the detection in mid-2014 of the terrestrial planet ($\sim$ 5 Earth masses) orbiting 
extremely old (10--12 Gyr) Kapteyn's star (Anglada-Escud{\'e} et al. 2014). The most remarkable thing is that this planet 
lies in the HZ. The star also has another super-Earth outside the HZ. 

\subsection{Potential habitability of old planetary systems}

The improved precision has also resulted in the rejection of three previously reported old planets HIP 13044 b and 
HIP 11952 b,c \citep[e.g.][]{set10, set12} as a genuine signal (Jones \& Jenkins 2014). However, it still leaves the 
number of old planets of at least 117 (92 hosts, see Table~\ref{table:all_old} in the Appendix) with 11 super-Earths 
(namely, Kepler-18 b; 55 Cnc e; Kapteyn's b,c; MOA-2007-BLG-192L b, OGLE-2005-BLG-390L b and five planets of Kepler-444) 
and all the rest gas giants, which do not fall into the category of habitable planets. However, because giant planets 
typically harbor multiple moons, the moons may be habitable and may even lie in the domain of a higher habitability, 
or even ``superhabitability" (Heller \& Armstrong 2014a). For example, \citet{schu} estimate the PHI for Jupiter to be 
only 0.4, while it is around 0.65 for Titan. There are 33 potentially habitable exomoons with habitable surfaces listed 
by HEC (excluding possibility of subsurface life), which have on average ESI higher than the potentially habitable planets. 
Heller et al. (2014) have
shown that the number of moons in the stellar HZ may even
outnumber planets in these circumstellar zones, and that massive exomoons are potentially detectable with 
current technology (Heller 2014).
Even though Population~II (Pop~II) stars are normally two order of magnitude less abundant in metals, they may harbor up 
to 10 potentially habitable rocky Earth-size subsolar objects each \citep{ssm}, either as planets or as moons orbiting 
gaseous giants. Planets can form at metallicities as low as $Z\sim 0.01~Z_\odot$ due to the centrifugal accumulation of 
dust \citep{ssm}. However, Pop~II stars could have formed in the metals-enriched pockets resulting from a non-perfect 
mixing in young galaxies when the Universe was as young as a few hundreds of Myr \citep{dy, vd}. They would be able to 
form planets in a traditional way, and our Galaxy may have a vast number of rocky planets residing in habitable zones. 
Such planets had longer time for developing biogenesis. Recently discovered 5 rocky planets orbiting 11.2 Gyr old star 
Kepler-444 (Campante et al. 2015) seems to confirm the previously suggested \citep{ssm} hypothesis. 

Direct measurements of metallicities and abundance pattern in the early Universe have recently become possible with
the discovery of extremely metal-poor (EMP) stars with metallicities as low as $10^{-5}$ of the solar value --- 
these objects are believed to represent the population next after the Population~III stars \citep{beer}. The relative 
abundances observed in the EMP stars are shown to stem from the explosions of Population~III intermediate-mass 
SNe with an enhanced explosion energy around $5\times 10^{51}$ erg \citep{jp}. These stars are also often found
to be overabundant in CNO elements. Interestingly, their relative abundance \citep{aoki,ito} is consistent with the abundance 
pattern of the Earth crust \citep{tay,yan} and the chemical composition of the human body \citep[see, e.g.][]{ni99}. 

Though Earth is rich in chemistry, living organisms use just a
few of the available elements: C, N, O, H, P and S, in biological
macromolecules: proteins, lipids and DNA, which can constitute up to 98\% of an organisms' mass (e.g., Alberts et al. 2002).
 Apart from hydrogen, these `biogenic' elements are all produced by the very first massive 
pop~III stars. Detection of substantial amount of CO and water in the spectrum of $z=6.149$ quasar SDSS J1148+5251 
shows, for example, that at $\sim 800$ Myr after the Big Bang, all the ingredients for our carbon-based life were already 
present. The initial episode of metal enrichment is believed to have occurred when the Universe was about 500--700 Ma --- the 
absorption spectra of high-redshift galaxies and quasars show significant amount of metals, in some cases up to 0.3 of the solar
 metallicity \citep[e.g.,][]{{sav},{nat}}. The abundance pattern of heavy elements in this initial enrichment contains a copious 
amount of elements sufficient for rocky planets to form within the whole range of masses \citep{br99,ab00,cl11,sta11}.

Therefore, planets formed in the early Universe and observed now as orbiting very 
old ($\gtrsim 9$ Gyr) Pop~II stars, may have developed and sustained life over the epochs when our Solar System 
had only started to form. In this way, the restricted use of six ‘biogenic’ elements may be considered as a
fossil record of an ancient life --- it is well known that at the molecular level, living organisms are strongly conservative. 
 The general 
direction of the biological evolution is in the increase of complexity of species rather than (chemical) diversity (Mani 1991). 
For example, paradoxically, both oxygen and water are destructive to all forms of carbon-based life (e.g. Bengston 1994). 
The presence of water reduces the chance of constructing nucleic acids and most other macromolecules 
(Schulze-Makuch \& Irwin, 2006). The toxic 
nature of oxygen necessitated the evolution of a complex respiratory metabolism, which again shows the strong chemical 
conservatism at the molecular level in that the living organisms developed the protection mechanisms to circumvent these 
problems rather than use other compounds.

\section{Observational Prospects}
\label{sec:prospects}

Recently, a 13.6 Gyr star was detected placing it as the oldest star in the Universe (SMSS J031300.36−670839.3,
Keller et al. 2014); the age was estimated by its metallicity [Fe/H]$\leq -7.41$. In spite of that, this star, 
believed to have formed from the remnants of the first-generation SN, was found
to contain carbon, metals like lithium, magnesium, calcium, and even methylidyne (CH). It is quite possible 
that such stars have planets that are directly observable in micron wavelength range. Such 
EMP stars are known to have low masses and, as such, the orbiting planets could be seen directly in the IR.  

The number of EMP stars is estimated to be around 250,000 within 500 pc in SDSS database \citep{aoki}, so the mean 
distance between them is about 10 pc. If each EMP star hosts an Earth-size planet, the flux from the planet 
at a distance $d$ in the IR range ($\lambda\sim 10~\mu$m) evaluated at the peak frequency (Wien's law) 
$\nu_{\rm T}=\a kT/h$, is 
\be
F^{\rm pl}_\nu =\s T^4_{\rm pl} \pi\left(\fr{R_{\rm pl}}{d}\right)^2\,.
\ee
We can rewrite this flux as
\be
F^{\rm pl}_\nu = 0.73\left(\fr{T_{\rm eq}}{300~{\rm K}}\right)^3\left(\fr{d}{10~{\rm pc}}
\right)^{-2}\left(\fr{R}{R_{_{\rm E}}}\right)^2~{\rm mJy}\,,
\ee
where $T_{\rm eq}$ is the equilibrium temperature of a planet and $R$ is its radius. 
For the Sun/Earth system, the ratio of the fluxes at a distance of 10 pc is 
\be
\fr{F_\nu^{\rm pl}}{F_{\ast}}= \fr{T_{\rm E}}{T_{\odot}}\left(\fr{R_{\rm E}}{R_{\odot}}\right)^2\sim 
4 \times 10^{-6}\,.
\ee
However, if we consider a super-Earth with $M \sim 5\,M_{\rm E}$,  $R\sim 2 R_{\rm E}$ and $T_{\rm eq}=300$ K, 
orbiting the star with $T=3000$ K and $R\sim 0.1\,R_{\odot}$ -- an M dwarf, we get an improvement of
\be\label{submm}
\fr{F_\nu^{\rm pl}}{F_{\ast}} =3.4 \times 10^{-3}\,.
\ee 
It seems challenging to detect such a weak contribution to a total flux from a planet even in the IR. There is, however, 
a possibility to distinguish the emission from the planet in IR molecular features, such as CH$_4$ or O$_2$, tracing either 
initiated biogenesis or developed metabolism. Detection of direct IR emission from O$_2$ on exoplanets going 
through the initial epoch of biogenesis, or which are already at a stage with developed biota, was discussed in 
\citet{chu} and \citet{rodl}, respectively. Rich IR to sub-mm spectra of methane (Niederer 2011, Hilico et al. 1987) also 
allow to optimistically view the future detection of this biosignature. Even at the low temperatures of EMP stars, $T_* \sim 
3000$ K, these molecules are unlikely to survive in sufficient amount in their atmospheres. Therefore, if such emission is observed 
from an EMP star, it should be considered as a direct indication of an orbiting rocky planet that has already entered the 
habitable epoch with growing $PHI$ (Eq.~\ref{eq:6}). The most promising way to identify habitable ({\it inhabited}) planets 
seems to look for simultaneous presence of water, O$_2$, O$_3$, CH$_4$ and N$_2$O in atmospheric spectra (e.g. Selsis et al. 
2002, Kiang et al. 2007, Kaltenegger et al. 2007). Though such observations can be used to detect planets with highly developed 
habitability orbiting old EMP stars, the expected fluxes in the IR are still below 
current sensitivity limits and might be only possible in the future. For example, the future {\it Millimetron} space observatory 
planned for launching in next decade (estimated launch 2025) will have the detection limit of 0.1~$\mu$Jy in 1 hour observation 
in 50-300 $\mu$m range (Kardashev et al 2014). A molecular CH$_4$ absorptions at $\sim 50~\mu$m can be detected by
{\it Millimetron} in 3 hours observations (Eq.~\ref{submm}) if a nearby (within 10 pc) habitable super-Earth planet transits an M-dwarf. 
  
\section{Summary}

\begin{itemize}

\item{} Age of a planet is an essential attribute of habitability along with such other factors as 
liquid water (or an equivalent solvent), rocky mantle, appropriate temperature, extended atmosphere, and so forth. 
The knowledge of the age of a ``habitable'' planet is an important factor in developing a strategy to 
search for complex (developed) life; 

\item{} Nearly half of the confirmed PHPs are young (with ages
less than $\sim 3.5$ Gyr) and may not have had enough time
for evolution of sufficiently complex life capable of
changing its environment on a planetary scale;

\item{} Planets do exist around old Pop~II stars, and recently discovered EMP stars (belonging presumably 
to an intermediate Pop~II.5) are good candidates for direct detection of orbiting planets in 
the IR and sub-mm wavelengths. Though currently only very few such potentially 
habitable planets are known, old giant planets may have habitable worlds in the form 
of orbiting moons; 

\item{} IR and sub-mm observations of terrestrial planets orbiting low-mass old stars is a promising way to 
trace biogenetic evolution on exoplanets in the solar neighbourhood. 

\end{itemize}

\section{Acknowledgments}

YS acknowledges the hospitality of RRI and IIA, Bangalore, when this work has been initiated. The authors 
thank Tarun Deep Saini for his useful comments, IIA Ph.~D. student A.~G.~Sreejith for help with graphics and 
IIA internship student Anuj Jaiswal for his contribution in the project of updating the HabCat. The authors also
thank the referees for their valuable comments which led
to considerable improvement in the paper.

This research has made use of the Extrasolar Planets 
Encyclopaedia at {\tt http://www.exoplanet.eu}, Exoplanets Data Explorer at {\tt http://exoplanets.org}, 
NASA Exoplanet Archive at {\tt http://exoplanetarchive.ipac.caltech.edu} and NASA Astrophysics Data System Abstract Service.

\onecolumn
\begingroup

\section*{Appendix~A. The list of stars with estimated ages $\geq 9$ Gyr.}

\begin{longtable}{lccl}
%\begin{center}
\caption{\small Stars with measured/estimated masses of $\geq 9$ Gy. Where possible, the mass of the planet is given, 
where the following abbreviations are used: J - Jupiter mass, S - Saturnian mass, N - Neptunian mass, E - 
Earth mass, sE - super-Earth. Along with the reference for the age, the method of determination is given, where possible.}\\
\hline \multicolumn{1}{l}{\textbf{Star name}} & \multicolumn{1}{c}{\textbf{Age}} &
\multicolumn{1}{c}{\textbf{Planets}}  & \multicolumn{1}{l}{\textbf{Refs., notes}}   \\ \hline 

\endfirsthead
\multicolumn{4}{c}%
{{\bfseries \tablename\ \thetable{} -- continued from previous page}} \\
\hline \multicolumn{1}{l}{\textbf{Star name}} &
\multicolumn{1}{c}{\textbf{Age}} &
\multicolumn{1}{c}{\textbf{Planets}}  & \multicolumn{1}{l}{\textbf{Refs., notes}}\\ \hline 
\endhead

\hline \multicolumn{4}{r}{{\bf Continued on next page}} \\ 
%\hline
\endfoot
%\hline
\endlastfoot
16CygB=HD217014  &  9-10   &    2.3J    & Saffe et al.'05, isoch, Li\\
GJ86=HR637       &  12.5    &    4J      &    -"-, Fe/H      \\
rho Crb=HD143761 &  11.9-12.1 &  1J      &    -"-,isoch, Fe/H \\
HD4208           &  12.4      &  0.8J    &    -"-, Fe/H      \\
HD16141=79Ceti   &  11.2      &  S=0.2J   &   -"-, isoch     \\
HD41004A         &  9.5       &  $>2.5$J    &    -"-, Fe/H     \\
HD45350          &  12.6      &  $\geq 2$J     &    -"-, isoch     \\
HD65216          &  10.2      &   $>1$J      &   -"-, Fe/H     \\
HD73526          &  10.3      & b$>2$J;c$>2.3$J  &   -"-, isoch     \\
HD76700          &  11.5      & hot $>0.1$J  &   -"-, isoch     \\
HD89307          &  12.2      & J          &   -"-, Fe/H     \\
HD108874         &  10.7-14.1 & b,c$>1$J    &   -"-, isoch     \\
HD114386         &  9.2       & 1.2J       &    -"-, Fe/H     \\
HD114729         &  11.9-12.5 & 1J         &    -"-, isoch     \\
HD134987         &  11.1      & b=0.8J;c=1.5J  &   -"-, isoch     \\
HD142022         &  9.4-17.2  & $>4.47$J     &    -"-, isoch     \\
HD154857         &  13.1      & 2 giants   &    -"-, Fe/H      \\
HD162020         &  9.5       & 14.4J       &    -"-, Fe/H    \\
HD168443         &  10.6      & b$>7.5$J;c$>17.5$J &  -"-, isoch     \\
HD168746         &  9.2-16    & gas giant   &  -"-,isoch, Fe/H \\ 
HD190228         &  12.5      &  giant      &  -"-, Fe/H      \\
HD195019         &  10.6      &  3.7J       &  -"-, isoch     \\
HD208487         &  10.8      &  1J         &  -"-, Fe/H      \\
HD216437         &  8.7       &  2J         &  -"-, isoch     \\
HD217107         &  9.5-9.9   &  1J         &                  \\   
HD181720=HIP95262  & 9.4-12.1  &  gas giant   & Santos et al. 2010  HAPRS\\
HD4308=HIP3497    & 11.5      &  N=14E     &    Haywood 2008  \\
HD6434=HIP5054    & 10.4      &  $>0.4$J     &    -"-     \\
HD37124=HIP26381  & 14.7      &  b,c,d$\geq 0.7$J &  -"-     \\
HD47536=HIP31688  & $9.33\pm 1.88$ &  b=5J,c=7J     &  -"-, Silva et al. 2006 \\
HD111232=HIP62534  &12.0     &   $>7$J         &   -"-     \\
HD114762=HIP64426  & 12.4    &   $>11$J        &   -"-    \\
Kapteyn's         & 10-12     &  b,c=sE+sE     &   -"-    \\    
PSR-B1620-26 (M4) & $12.8\pm 2.6$ &  2.5J       &     SSM     \\
BD20-2457         & 12.7       &  b=12.47J,c=21.42J &  SSM      \\
HD155358          & 11.9      &  b=0.85J,c=0.82J    &   SSM     \\
HAT-P-26          &  9        &  8J            &  exoplanets.eu\\
HD102365          &  9         &  N=16E        &        -"- \\
HD96063           &  9         &  0.9J         &        -"-  \\
HD103197          & 9.1      &  gas giant, 31E  &     -"-     \\
HD154672          & 9.28      &   5J            &       -"- \\
KOI-1257a         & 9.3       &  1.45J          &      -"- \\
HD47536           & 9.33      &  b=5J,c=7J          &     -"-  \\
HD4203            & 9.41      &  b=1.2J,c=2J        &    -"-   \\
42Dra=hd170693    & 9.49      &   $\sim4$J           &           \\
HD11964           & 9.56      &  b=0.1J(N),c=0.6J  &  -"-   \\
HD88133           & 9.56      &   $>0.3$J          &    -"-   \\
HATS-2            & 9.7       &    1.3J          &  -"-   \\
HD87883           & 9.8       &    $>1.8$J        &    -"-   \\        
Kepler-46         & 9.9      &    b=S,c=0.37J        &  -"-   \\
Kepler-18         & 10       &  6.9E=sE;17E=N;16E=N&  -"-   \\
V391Peg           & 10       &    $>3.2$J        &     -"-   \\
HAT-P-38          & 10.1     &    0.27J=S      &  -"-   \\
55Cnc=HD75732     & 10.2    &  0.8J=S,$>0.17$J=S,3.8J,8.63E=sE,$>0.155$J=S &   \\ 
HAT-P-21         &  10.2     &   4J          &  -"-   \\
HD109749         &  10.3     &    $\sim 0.3$J=S       &  -"-   \\
Kepler-10        &  10.6     &   b=3.3J,c=17J          &  -"-   \\
CoRoT-17         &   10.7    &  hot 2.4J &  -"-   \\
CoRoT-24         &   11      & b,c=N,N     &  -"-   \\
WASP-37          &   11      &    1.8J        &  -"-   \\
WASP-6           &   11      &    0.88J      &  -"-   \\
WASP-11-HAT-P-10 &  11.2     &     0.8J      &  -"-   \\
WASP-19          &   11.5    &    $\sim 1.1$J        &  -"-   \\
WASP-97          &   11.9     &    hot 1.3J &  -"-   \\
HD152581         &   12       &     1.5J      &  -"-   \\
HD190360=Gl777a  &   12.11    &    0.06J=N,$\sim 1.6$J  &  -"-   \\
HD99109          &   12.2     &    0.5J          &  -"-   \\
HAT-P-18         &   12.4     &     0.1J=S      &  -"-   \\
HAT-P-22         &   12.4     &    2J        &  -"-   \\
PSR1719-14       &   12.5     &    $\sim 1$J        &  -"-   \\
HD164922         &   13.4     &    $\sim 0.4$J=S    &  -"-   \\
WASP-29          &    15      &    0.25J=S     &  -"-   \\
rho Indus        &   12.959    &     2J & exoplanet.eu + exoplanets.org\\
SAO38269=BD=48738  & 12.217     &   $\sim 1$J     &     -"- \\
OGLE2005-BLG-071L  &  11.404    &   3J       &    -"- \\
DP Leonis          &   11.23    &      6J   &  -"-   \\
HD37605            &   10.712   &    b,c=J,J   &  -"-   \\
OGLE2003-BLG-235L  &  10.471    &    2J     &  -"-   \\
MOA2007-BLG-192L   &   10.42    &    3E=sE    &  -"-   \\
MOA2009-BLG-387L   &   10.266   &    $\sim 3$J      &  -"-   \\
NN Serpent's       &    10.153  &   b=7J,c=2J     &  -"-   \\
iota Draco=12 Dra   &    10.015  &    12J      &  -"-   \\
Gliese 649          &    9.998   &     1S      &  -"-   \\
18 Delphini         &   9.897    &      10J     &  -"-   \\
OGLE2005-BLG-390L   & 9.587      &     5.5E=sE   &  -"-   \\
WASP-5              &    9.582   &       1.6J     &  -"-   \\
Gliese 253          &  9.451     &       b,c=N,N   &  -"-   \\
HD1690              &  9.332     &       6J        &  -"-   \\
WASP-33=HD15082     &    9.106   &        $\sim 4$J      &  -"-   \\
WASP-23             &  9.033     &      $\sim 1J$      &  -"-   \\
OGLE2005-BLG-169L    &  9.623     &        1U     &  -"-   \\
Kepler-108=KOI119.02 & $ 8.9\pm 3.7 $ &    8E    &   NASA Exoplanet archive \\
Kepler-444                   &$ 11.23\pm 0.99 $ &  b,c,d,e,f -- all $<$ Venus & Campante et al. 2015, astroseismology \\ 
\hline
%\end{tabular}
%\end{center}
\label{table:all_old}
\end{longtable}
\endgroup

\newpage
\section*{Appendix~B. The ‘‘ Near 100 ’’ ---  a subset from the nearest 100 star systems.}
\vskip -0.3in
\begin{table}[ht!]
\begin{center}
\caption{\small Selection from the ``Near 100" of the stars with estimated ages. Information on planetary systems is 
important in devising future space missions focussing on astrobiology, thus a column is included whether a particular 
star has a planet or not, with planetary mass(es).}
\begin{tabular}{llll}
\hline
Star	Age (Gyr)& 	Planets	 & 	Planet(s)    	 & Notes             \\\hline
GJ 338AB & 0.025-0.3	 & 	No		                          & 	                    \\
GJ 1	 & $(0.1\pm 0.1)\times 10^{-3}$ &	No	&	                          \\
GJ 406	 & 0.1-0.35		 & 	No				 &	                                 \\
GJ 873	 & 0.1-0.9		 & 	No				 &	                                 \\
GJ 876A	 & 0.1-5.0		 & 	Yes, 4	  &$b=2.2756\pm 0.0045$J,c=0.714J,d=0.0215J,e=0.046J; 2 in HZ \\
GJ 1111	 & 0.2		       & 	No				 &	                                 \\
GJ 244A	 & 0.2-0.3		 & 	No				 &	                                  \\
GJ 244B	 & 0.2-0.3		 & 	No				 &	                                 \\
GJ 144	 & 0.2-0.8		 & 	Yes, 1			 &$(1.55\pm 0.24)$J		 \\
GJ 566A	 & 0.2		        & 	No				 &	                                 \\
GJ 65A	 & $< 1$		 & 	No				 &	                                 \\
GJ 65B	 & $< 1$	       & 	No				 &	                                 \\
GJ 729	 & $< 1$		 & 	No				 &	                                 \\
GJ 768	 & $< 1$		 & 	No				 &	                                 \\
GJ 881   & $0.4\pm 0.04$         & Yes, 1   &  $< 2-3$J                                   \\
GJ 674   & $0.55 \pm 0.45$       & Yes, 1   & $\geq 11.8$E                                             \\
GJ 176   & 0.56                  & Yes, 1    &$>8.4$E                     \\
GJ 663A	 & $0.6-1.8$		 &	No				 &	                                 \\
GJ 280B  & 1.37		 &	No		              &	                                 \\
GJ 440	 & 1.44		       &	No		              &			                    \\
GJ 702A	 & 1.9		       &   No		              &	                                 \\
GJ 667C  &2-10, 2           &  Yes, 2 & b$\geq 5.661\pm 0.437$E,c$\geq3.709\pm 0.682$E; 1 in HZ                \\
GJ 876   & $2.5\pm 2.4$ pop.~I &Yes, 4 & b=$2.2756\pm 0.0045$J,c=$0.7142\pm 0.0039$J,d=$6.83\pm 0.4$E,e=$14.6\pm 1.7$E \\ 
GJ 280A	 & 3 		       &	No				 &	                                 \\
GJ 849   & $> 3$middle age dwarf& Yes, 2    &  b=$0.90\pm 0.04$J,c=0.77J                                          \\
GJ 35	 & 4 		       &	No	 &	                                                      \\
GJ 764	 & 4.7	             &   No				 &	                                 \\
GJ 551	 & 4.85		       &	No				 &	                                 \\
GJ 442A  & 4.5-5.7            & Yes, 1                           & $0.05\pm 0.008$J                    \\
GJ 71	 & 5.8	             &    Yes, 5  &b=$2.0\pm 0.8$E,c=$3.1\pm 1.4$E,d=$3.6\pm 1.7$E,e=$4.3\pm 2.01$E,f=$0.783\pm 0.012$E	  \\
GJ 411	 & 5.0-10.0     	& No				 &	                                 \\
GJ 559A	 & 5.0-7.0	       & No				 &	                                 \\
GJ 559B	 & 5.0-7.0	       & Yes, 1			 &1.13E, Outside HZ\\
GJ 34A	 &$ 5.4\pm 0.9$	&No				       &	                                 \\
GJ 1221	 & 5.69       		 &	No				 &	                                 \\
GJ 820A	 & $6.1\pm 1$     &	No				 &	                                 \\
GJ 820B	 &$ 6.1\pm 1$     &   No				 &	                                 \\
GJ 139	 & 6.1-12.7	       & Yes, 3 sE			 &b$\geq 2.7\pm 0.3$E,c$\geq 2.4\pm 0.4$,d$\geq 4.8\pm 0.6$E\\ 
GJ 506   & 6.1-6.6               & Yes, 3 & b=$5.3\pm 0.5$E, c=$18.8\pm 1.1$E,d=$23.7\pm 2.7$E              \\ 
GJ 380	 & 6.6			 &No				 &	                                 \\
GJ 780	 & 6.6-6.9		 &	No				 &Best SETI Target acc. To Turnbull\&Tarter (2003)\\
GJ 223.2 & 	$6.82\pm 0.02$ &	No			 &		                              \\
GJ 785   & 7.5-8.9              & Yes, 2   & b$\geq 16.9\pm 0.9$E,c$\geq 24\pm 5$E                            \\
GJ 783A	 & 7.7		        & 	No				 &	                                    \\
GJ 581	 & $8\pm 1$	        &Yes, 3		 &e$>1.7\pm 0.2$E,b$>15.8\pm 0.3$E,c$>5.5\pm 0.3$E; 2 in HZ	           \\
GJ 832	 & 9.24			 &Yes, 2	 &b$\geq 0.64\pm 0.06)$J,c$\geq 5.4\pm 1$E	     \\
GJ 191	 & 10			 &Yes, 2		            &	$(4.8\pm 1)$E	            \\
GJ 699	 & $\sim 10$		 &	No				 &	                                  \\
GJ 892	 & 12.46	         &No				 &	                                 \\\hline
\end{tabular}
\end{center}
\label{table:habcat}
\vskip -0.1in
{\it Note}: Some of the catalogues used in the study: The Open Exoplanet Catalogue; Extrasolar Planets Encyclopaedia; 
NASA Exoplanet Archive; Exoplanets Data Explorer.
\end{table}
\end{document}